\pdfoutput=1
\documentclass[sigplan,screen,rgb]{acmart}
\usepackage{pifont}
\newcommand{\Checkmark}{\ding{51}}
\newcommand{\XSolidBrush}{\ding{55}}

\startPage{1}

\setcopyright{none}

\usepackage{booktabs}   
\usepackage{subcaption} 

\usepackage{fancyvrb}

\newcommand{\trec}[1]{\ensuremath{\{#1\}}}
\newcommand{\ttensor}[2]{\ensuremath{[{#1} \text{ \texttt{of} } {#2}]}}
\newcommand{\ttop}{\ensuremath{\top}}

\newcommand{\tool}{{\em Ariadne}}

\setcopyright{acmcopyright}
\acmPrice{15.00}
\acmDOI{10.1145/3211346.3211349}
\acmYear{2018}
\copyrightyear{2018}
\acmISBN{978-1-4503-5834-7/18/06}
\acmConference[MAPL'18]{2nd ACM SIGPLAN International Workshop on Machine Learning and Programming Languages}{June 18, 2018}{Philadelphia, PA, USA}

\begin{document}

\title[\tool]{\tool: Analysis for Machine Learning Programs}         


\author{Julian Dolby}
\author{Avraham Shinnar}
\author{Allison Allain}
\author{Jenna Reinen}
\email{{dolby, shinnar, acallain}@us.ibm.com, Jenna.Reinen@ibm.com}
\affiliation{
  \institution{IBM Research}           
  \streetaddress{1101 Kitchawan Road}
  \city{Yorktown Heights}
  \state{NY}
  \postcode{10598}
  \country{United States}                   
}

\begin{abstract}
  Machine learning has transformed domains like vision and
  translation, and is now increasingly used in science, where the
  correctness of such code is vital.  Python is popular for machine learning, in part
  because of its wealth of machine learning libraries, and is
  felt to make development faster; however, this dynamic language has
  less support for error detection at code creation time than tools
  like Eclipse.  This is especially problematic for
  machine learning: given its statistical nature, code with subtle
  errors may run and produce results that look 
  plausible but are meaningless.  This can vitiate scientific
  results.  We report on \tool: applying a static
  framework, WALA, to machine learning code that uses TensorFlow.  We
  have created static analysis for Python, a type
  system for tracking tensors---Tensorflow's core data
  structures---and a data flow analysis to track their usage.  We
  report on how it was built and present some early results. 
\end{abstract}



\maketitle

From translation\cite{DBLP:journals/corr/JohnsonSLKWCTVW16} to vision\cite{Prince:2012:CVM:2344089}, machine learning has transformed many domains, and is increasingly used in science.  For example, in clinical and cognitive neuroscience, classifiers have been used as key evidence in identifying disease\cite{Sun2009, Reinen2018}, and to further scientific understanding of cognitive\cite{Wager2013} and mental states~\cite{POLDRACK2008223} from complex patterns of brain activation.  It is vital that underlying code be correct.

 In most programming domains, preventing and catching programming errors is aided by supportive tooling. Tools also aid programming by suggesting appropriate completions while writing code, and refactoring code\cite{1999:RID:311424} to improve design and reuse.  These tools usually are part of IDEs, of which Eclipse and IntelliJ are well known for Java. 

 Machine learning code is commonly written in Python, in part because of its wealth of machine learning libraries, but also reflecting a more-general trend toward dynamic programming languages. For example, witness the dominance of JavaScript in Web development\cite{JSShare}.  While these languages are often felt to make development faster, the price is having less error detection at code creation time.

 This is especially problematic for machine learning: due to its statistical nature, code with subtle errors may run and produce results that look plausible but are meaningless.  Meaningless results can occur when logical errors cause code to find spurious correlations that arise in big data.  In science, this could appear to support erroneous hypotheses.

 Supportive tooling could ameliorate such problems, but it is less sophisticated for dynamic languages, in large part because such tooling is usually driven by static types that these languages lack.  However, without static types, information can still be gleaned using static analysis; this has been done in tools like IBM's AppScan\cite{AppScan} security analysis for JavaScript.

 In this paper, we apply static analysis to machine learning code that uses Tensorflow\cite{tensorflow2015-whitepaper}.  We extend WALA to support Python, build a type system for tracking tensors---one of machine learning's core data structures---and an analysis of tensor usage.  Our contributions are the following:
\begin{enumerate}
\item Application of WALA program analysis to Python
\item Type system capturing semantic properties of tensors
\item An open-source implementation
\end{enumerate}

We start with a running example in Section~\ref{sec:example}, and then discuss how we analyze the code in Section~\ref{sec:wala}.  We define our type system for tensors in Section~\ref{sec:types}, and present how it is used to analyze an example of tensor usage in Section~\ref{sec:dataflow}.  We then survey related work and conclude.

\section{Example}
\label{sec:example}

\newcommand{\tlabel}[1]{\textit{#1}}
\newcommand{\tfield}[1]{\textsf{#1}}

\begin{figure*}[p]
  \centering
 {\scriptsize
\begin{Verbatim}[commandchars=\@\|\^,numbers=left,numbersep=6pt]
# Import MNIST data
from tensorflow.examples.tutorials.mnist import input_data
mnist = input_data.read_data_sets("/tmp/data/", one_hot=False)	@label|mnist_read^		## {@tfield|train^:@trec|@tfield|images^:@ttensor|@tlabel|@tlabel|batch^^,@tlabel|y^(28)*@tlabel|x^(28)^|@tlabel|channel^^^,
import tensorflow as tf								  # @tfield|test^:@trec|@tfield|images^:@ttensor|@tlabel|@tlabel|batch^^,@tlabel|y^(28)*@tlabel|x^(28)^|@tlabel|channel^^^}
# Training Parameters
learning_rate = 0.001
num_steps = 2000
batch_size = 128
# Network Parameters
num_input = 784 # MNIST data input (img shape: 28*28)
num_classes = 10 # MNIST total classes (0-9 digits)
dropout = 0.75 # Dropout, probability to keep units
# Create the neural network
def conv_net(x_dict, n_classes, dropout, reuse, is_training): @label|def_conv_net^ ## @ensuremath|(@tfield|x_dict^:@trec|@tfield|images^:@ttensor|@tlabel|@tlabel|batch^^,@tlabel|y^(28)*@tlabel|x^(28)^|@tlabel|channel^^^)@to@ttensor|@tlabel|batch^,@tlabel|n_classes^^|@tlabel|channel^^^
    # Define a scope for reusing the variables
    with tf.variable_scope('ConvNet', reuse=reuse):
        # TF Estimator input is a dict, in case of multiple inputs
        x = x_dict['images'] @label|heap_read_images^							 ## @ttensor|@tlabel|@tlabel|batch^^,@tlabel|y^(28)*@tlabel|x^(28)^|@tlabel|channel^^
        # MNIST data input is a 1-D vector of 784 features (28*28 pixels)
        # Reshape to match picture format [Height x Width x Channel]
        # Tensor input become 4-D: [Batch Size, Height, Width, Channel]
        x = tf.reshape(x, shape=[-1, 28, 28, 1]) @label|reshape^				     ## @ttensor|@tlabel|@tlabel|batch^^,@tlabel|y^(28),@tlabel|x^(28),1^|@tlabel|channel^^
        # Convolution Layer with 32 filters and a kernel size of 5
        conv1 = tf.layers.conv2d(x, 32, 5, activation=tf.nn.relu)@label|conv1^		     ## @ttensor|@tlabel|@tlabel|batch^^,@tlabel|y^(28),@tlabel|x^(28),1^|@tlabel|channel^^
        # Max Pooling (down-sampling) with strides of 2 and kernel size of 2		
        conv1 = tf.layers.max_pooling2d(conv1, 2, 2)@label|pool1^				  ## @ttensor|@tlabel|batch^,@tlabel|y^(14),@tlabel|x^(14),1^|@tlabel|channel^^
        # Convolution Layer with 64 filters and a kernel size of 3
        conv2 = tf.layers.conv2d(conv1, 64, 3, activation=tf.nn.relu)		 ## @ttensor|@tlabel|@tlabel|batch^^,@tlabel|y^(14),@tlabel|x^(14),1^|@tlabel|channel^^
        # Max Pooling (down-sampling) with strides of 2 and kernel size of 2
        conv2 = tf.layers.max_pooling2d(conv2, 2, 2)@label|pool2^				  ## @ttensor|@tlabel|@tlabel|batch^^,@tlabel|y^(7),@tlabel|x^(7),1^|@tlabel|channel^^
        # Flatten the data to a 1-D vector for the fully connected layer
        fc1 = tf.contrib.layers.flatten(conv2)@label|flatten1^					## @ttensor|@tlabel|batch^,@tlabel|y^(7)*@tlabel|x^(7)^|@tlabel|channel^^
        # Fully connected layer (in tf contrib folder for now)
        fc1 = tf.layers.dense(fc1, 1024)					      ## @ttensor|@tlabel|batch^,1024^|@tlabel|channel^^
        # Apply Dropout (if is_training is False, dropout is not applied)
        fc1 = tf.layers.dropout(fc1, rate=dropout, training=is_training)	      ## @ttensor|@tlabel|batch^,1024^|@tlabel|channel^^
        # Output layer, class prediction
        out = tf.layers.dense(fc1, n_classes)					 ## @ttensor|@tlabel|batch^,@tlabel|n_classes^^|@tlabel|channel^^
    return out

# Define the model function (following TF Estimator Template)
def model_fn(features, labels, mode): @label|def_model_fn^		 		## (@tfield|features^:{@tfield|images^:@ttensor|@tlabel|batch^,@tlabel|y^(28)*@tlabel|x^(28)^|@tlabel|channel^^},@tfield|labels^:@ttensor|10^|@tlabel|label^^)@ensuremath|@to@ttop^
    # Build the neural network
    # Because Dropout have different behavior at training and prediction time, we
    # need to create 2 distinct computation graphs that still share the same weights.
    logits_train = conv_net(features, num_classes, dropout, reuse=False, is_training=True) @label|call_conv_net^ 	## @ttensor|@tlabel|batch^,@tlabel|10^^|@tlabel|channel^^
    logits_test = conv_net(features, num_classes, dropout, reuse=True, is_training=False)	   ## @ttensor|@tlabel|batch^,@tlabel|10^^|@tlabel|channel^^
    # Predictions
    pred_classes = tf.argmax(logits_test, axis=1)
    pred_probas = tf.nn.softmax(logits_test)
    # If prediction mode, early return
    if mode == tf.estimator.ModeKeys.PREDICT:
        return tf.estimator.EstimatorSpec(mode, predictions=pred_classes)
    # Define loss and optimizer
    loss_op = tf.reduce_mean(tf.nn.sparse_softmax_cross_entropy_with_logits(logits=logits_train, labels=tf.cast(labels, dtype=tf.int32)))
    optimizer = tf.train.AdamOptimizer(learning_rate=learning_rate)
    train_op = optimizer.minimize(loss_op, global_step=tf.train.get_global_step())
    # Evaluate the accuracy of the model
    acc_op = tf.metrics.accuracy(labels=labels, predictions=pred_classes)
    # TF Estimators requires to return a EstimatorSpec, that specify
    # the different ops for training, evaluating, ...
    estim_specs = tf.estimator.EstimatorSpec( 
        mode=mode,
        predictions=pred_classes,
        loss=loss_op,
        train_op=train_op,
        eval_metric_ops={'accuracy': acc_op})
    return estim_specs

# Build the Estimator
model = tf.estimator.Estimator(model_fn) @label|store_model_fn^ @label|create_estimator^
# Define the input function for training
input_fn = tf.estimator.inputs.numpy_input_fn( 
    x={'images': mnist.train.images}, @label|heap_write_images^ y=mnist.train.labels,
    batch_size=batch_size, num_epochs=None, shuffle=True)
# Train the Model
model.train(input_fn, steps=num_steps) @label|indirect_call_model_fn^
\end{Verbatim}
}
\caption{Example using MNIST data}
\label{mnist-example}
\end{figure*}


Figure~\ref{mnist-example}~\cite{mnist-example-url} is an example that
uses well-known MNIST~\cite{MNIST} data to train image recognition.
It reads training data on line~\ref{mnist_read}.  To illustrate the
kind of checking we want, we focus on line~\ref{reshape}, which is a
{\tt reshape} operation, along with a comment describing it. The
comment is detailed, describing aspects of the original MNIST data and
the input data format expected by the subsequent {\tt conv2d}
operation.  The {\tt reshape} operation transforms a tensor, using a new shape to access the same underlying data.  A {\tt reshape} can transform dimensions, but the total size must be the same since it is accessing the same underlying data.  Thus, this piece of code depends on the structure of the MNIST data, which is only documented in comments; hence, there is nothing in the program to help the programmer get it right.

Our hypothesis is that operations such as this one could be checked for correctness, and perhaps even generated automatically, using type system and program analysis technology and minimal declarations by the programmer to specify input formats. This would check correctness in terms of both dimensions and types of data. If the types of MNIST and its dimensions were declared when the data is read, this information could be used to verify the reshape.

Assume that MNIST, at line~\ref{mnist_read}, were declared to be of a
type indicating the semantics of its dimensions, $$\ttensor{batch, x(28)*y(28)}{channel}$$
in our syntax; note that $batch$, $x$ and $y$ have no particular
meaning, but are simply names meaningful to the user for
describing the tensor dimensions. The declaration of the second
dimension specifies it to be composed from two logical dimensions, $x$
and $y$. Further assume that type inference could track that
declaration to line~\ref{heap_read_images}, and we had a declaration
for what {\tt conv2d} expects. In that case, we would be able to
observe that the reshape does the right thing, even observing that
28*28*1=784, so those dimensions evenly match the two logical
dimensions of the second dimension in MNIST. We would end up with a
type like $$\ttensor{batch, x(28), y(28), 1}{channel}$$
where the $x$, $y$ and $channel$ come from the conv2d operation.


\section{WALA}
\label{sec:wala}

 Existing program analysis frameworks
 (e.g. Doop\cite{Bravenboer:2009:SDS:1640089.1640108}, 
Soot\cite{Vallee-Rai:1999:SJB:781995.782008},
Safe\cite{SAFE_FOOL_2012}, WALA\cite{WALA}) provide the whole-program
analysis that would enable analyzing examples like in
Section~\ref{sec:example}; they can build call graphs describing how
functions interact and analysis of how objects behave.  But none of
these, to our knowledge, support Python. 
Fortunately, WALA does have a front end designed to make
adding new languages as easy as possible.  WALA supports Java and
JavaScript, and has been shown to scale on substantial Java
enterprise~\cite{techrep:DOMO} applications.  The
Java applications also demonstrated 
thoroughgoing support for modeling frameworks\cite{Sridharan:2011:FTA:2048066.2048145}, vital for the many
frameworks employed in Java enterprise code.  The JavaScript
applications showed that WALA can be applied to dynamic languages and
get results on security analysis\cite{Guarnieri:2011:SWW:2001420.2001442}.  After highlighting the challenges
faced by analysis, we describe how WALA allows us to model Python.

\subsection{Challenges}
In general, machine learning code uses all the normal features of programming languages, such as classes, methods, heap state; hence, program analysis must handle them.  This requires the kind of program analysis that has been used on a wide range of programming languages.  Three necessary aspects are the following, illustrated by Figure~\ref{mnist-example}:
\begin{description}
\item[Call graph construction] is required even to understand calls like the one on line~\ref{call_conv_net} to the function {\tt conv\_net} at line~\ref{def_conv_net}.  This is a direct call that could be handled easily, but there are more-complex cases that require higher-order functions.  The call to {\tt model\_fn} at line ~\ref{def_model_fn} is indirect from line~\ref{indirect_call_model_fn}, necessitating analysis of first-class functions; the function is stored by the constructor at line~\ref{store_model_fn} and that callback is ultimately invoked by the call at line~\ref{indirect_call_model_fn}.
\item[Pointer analysis] is needed to know that the read of the $images$ property at line~\ref{heap_read_images} is reading what is written by the store of $images$ at line~\ref{heap_write_images}; it requires pointer analysis of some kind to understand that the objects at those two points might be aliased.  Unlike Java, there are not even any types to use for a coarse approximation, and the same goes for functions, as well.  The function called indirectly at line~\ref{indirect_call_model_fn} is written at line~\ref{store_model_fn}, and the connection is through state in the heap.
\item[Library modeling] is what makes the connection between the call at line~\ref{store_model_fn} and the corresponding call to that function at line~\ref{indirect_call_model_fn}.  There is no such connection in the source code being analyzed, and the tensorflow source is not typically available when analyzing a program, as is common with other frameworks, especially in the enterprise space.  Analysis has used models for frameworks such as J2EE and Apache Struts, and this rich modeling machinery can be reused to model how Tensorflow behaves with respect to an application.
\end{description}
The pieces are all connected, with pointer analysis being needed to understand function pointers, which is in turn needed to include them in the call graph, which in turn adds new code with pointers to analyze.  It all is supported by rich models of the library.  Fortunately, these challenges are similar to what analysis frameworks have long had to do for other languages, especially Java and JavaScript, and we are able to leverage that technology.  

\subsection{Front End}
\label{front-end}

 WALA provides a framework for handling new languages, the Common
 Abstract Syntax Tree (CAst) system.  The supported way to handle new
 languages is to translate a normal language AST into CAst, and use
 built in support for translating CAst into WALA's Internal
 Representation (IR) for analysis; this IR is a fairly traditional
 three-address code in Static Single Assignment (SSA) form.  We use the existing Jython
 functionality to create Python ASTs for translation to CAst.  WALA
 translation support takes three major forms; see the classes
 (\url{https://github.com/wala/WALA}) and JavaDoc
 (\url{https://github.io/wala/javadoc}) of WALA for full details.  We
 focus here on aspects most relevant to our translation; we illustrate
 with concrete examples showing how Python is translated using methods
 from a visitor class over the Python AST, provided by Jython.  The
 full translation code is at \url{https://github.com/wala/ML}. 

\begin{itemize}
\item Some constructs are standard enough across languages that CAst
  provides a construct with translation to IR; e.g., {\tt if}, {\tt while}, {\tt goto} (including {\tt break} and {\tt continue}).  For such constructs, simply creating the appropriate CAst construct suffices.  Figure~\ref{cast-normal} shows translation of an {\tt if} statement; many forms such as assignments, arithmetic operations and constants are similar.

\begin{figure*}[p]
\centering
 {\scriptsize
\begin{Verbatim}[commandchars=\#\|\^,numbers=left,numbersep=6pt]
public CAstNode visitIf(If arg0) throws Exception {
  return Ast.makeNode(CAstNode.IF_STMT,
    arg0.getInternalTest().accept(this),
    block(arg0.getInternalBody()),
    block(arg0.getInternalOrelse()));
}
\end{Verbatim}
}
\caption{CAst generation for ordinary constructs.  {\tt Ast} is a factory for CAst nodes, and {\tt CAstNode} holds constants for different kinds of nodes.  A Python {\tt if} statement, {\tt arg0}, is being translated, with accessors for its test ({\tt getInternalTest}), then block ({\tt getInternalBody}) and else block ({\tt getInternalOrelse}).  The {\tt accept} method invokes the visitor recursively on children; the {\tt block} method is a helper that does the same for multiple statements.}
\label{cast-normal}
\end{figure*}

\item Some common constructs differ significantly in details; e.g. field accesses, where notion of what constitutes a field name vary.  Java has fixed constants, whereas dynamic languages like JavaScript and Python allow variables to name fields as well.  For such constructs, WALA provides CAst constructs with hooks in the IR translator to specialize it.  We illustrate the two pieces with CAst construction in Figure~\ref{cast-customizable}. and then the hook in IR translation from CAst in Figure~\ref{ir-customizable}.

\begin{figure*}[p]
\centering
{\scriptsize
\begin{Verbatim}[commandchars=\@\|\^,numbers=left,numbersep=6pt]
public CAstNode visitSubscript(Subscript arg0) throws Exception {
  return notePosition(Ast.makeNode(CAstNode.OBJECT_REF, 
    notePosition(arg0.getInternalValue().accept(this), arg0.getInternalValue()), 
    notePosition(arg0.getInternalSlice().accept(this), arg0.getInternalSlice())), arg0);
}
\end{Verbatim}
}
\caption{CAst generation for customizable constructs.  {\tt Ast} is a factory for CAst nodes, and {\tt CAstNode} holds constants for different kinds of nodes.  A Python field access, {\tt arg0}, is being translated, with accessors for the object ({\tt getInternalValue}) and field ({\tt getInternalSlice}).  The access could be a read or write.  {\tt notePosition} records source locations.}  
\label{cast-customizable}
\end{figure*}

\begin{figure*}[p]
\centering
{\scriptsize
\begin{Verbatim}[commandchars=\@\|\^,numbers=left,numbersep=6pt]
protected void doFieldRead(WalkContext context, int result, int receiver, CAstNode elt, CAstNode parent) {
  if (elt.getKind() == CAstNode.CONSTANT && elt.getValue() instanceof String) {
    FieldReference f = FieldReference.findOrCreate(PythonTypes.Root, Atom.findOrCreateUnicodeAtom((String)elt.getValue()), PythonTypes.Root);
    context.cfg().addInstruction(Python.instructionFactory().GetInstruction(context.cfg().getCurrentInstruction(), result, receiver, f)); @label|add_read_inst^
  } else ...
}
\end{Verbatim}
}
\caption{IR generation for customizable constructs.  IR generation for a field read, showing the case where the field name is a constant.  {\tt result} is the value number for the result, {\tt receiver} denotes the object, {\tt elt} is the CAst node of the field.  {\tt getValue} retrieves the constant value of a node; {\tt findOrCreate} makes an IR field reference for the field.  Line~\ref{add_read_inst} creates a field get instruction and adds it to the IR.}
\label{ir-customizable}
\end{figure*}

\item Idiosyncratic constructs are easiest to model in a custom way.
  For this, WALA provides the ability to add new CAst node types and
  also the generic {\tt primitive} CAst node type with custom
  translation to IR.  For example, we currently model the {\tt import}
  construct of Python as a static function call.  We illustrate with
  CAst generation of {\tt import} as {\tt primitive} in
  Figure~\ref{cast-idiosyncratic}, and show IR translation in Figure~\ref{ir-idiosyncratic}.

\begin{figure*}[p]
\centering
{\scriptsize
\begin{Verbatim}[commandchars=\@\|\^,numbers=left,numbersep=6pt]
public CAstNode visitImport(Import arg0) throws Exception {
 int i = 0;
 CAstNode[] elts = new CAstNode[ arg0.getInternalNames().size()*2 ];
 for(alias n : arg0.getInternalNames()) {
  elts[i++] = Ast.makeNode(CAstNode.DECL_STMT,
   Ast.makeConstant(new CAstSymbolImpl(name(n), PythonCAstToIRTranslator.Any)));
  elts[i++] = Ast.makeNode(CAstNode.ASSIGN,
   Ast.makeNode(CAstNode.VAR, Ast.makeConstant(name(n))),
   Ast.makeNode(CAstNode.PRIMITIVE, Ast.makeConstant("import"),   Ast.makeConstant(n.getInternalName().replaceAll("[.]", "/"))));
 }
 return Ast.makeNode(CAstNode.BLOCK_STMT, elts);
}
\end{Verbatim}
}
\caption{CAst generation for idiosyncratic constructs.  {\tt Ast} is a factory for CAst nodes, and {\tt CAstNode} holds constants for different kinds of nodes.  The Jython AST has multiple imports as one node, so this code loops over them, making each import the result of a {\tt primitive} {\tt import}, with the first argument being the module name.  The {\tt DECL\_STMT} create a new variable, and the {\tt ASSIGN} sets its value. These statements are grouped together into a block with a {\tt BLOCK\_STMT}.}
\label{cast-idiosyncratic}
\end{figure*}

\begin{figure*}[p]
\centering
 {\scriptsize
\begin{Verbatim}[commandchars=\@\|\^,numbers=left,numbersep=6pt]
protected void doPrimitive(int resultVal, WalkContext context, CAstNode primitiveCall) {
 if (primitiveCall.getChildCount() == 2 && "import".equals(primitiveCall.getChild(0).getValue())) {
  TypeReference imprt = TypeReference.findOrCreate(PythonTypes.pythonLoader, "L" + primitiveCall.getChild(1).getValue());
  MethodReference call = MethodReference.findOrCreate(imprt, "import", "()L" + primitiveCall.getChild(1).getValue());
  int idx = context.cfg().getCurrentInstruction();
  context.cfg().addInstruction( @label|module_call_start^
   Python.instructionFactory().InvokeInstruction(
    idx, resultVal, new int[0], context.currentScope().allocateTempValue(),  CallSiteReference.make(idx, call, Dispatch.STATIC), null)); @label|module_call_end^
 }
}
\end{Verbatim}
}
\caption{IR generation for idiosyncratic constructs.  IR translation of imports as function calls.  {\tt TypeReference} and {\tt MethodReference} create a reference to method corresponding to a module (only built-in modules are handled for now). Lines~\ref{module_call_start} to~\ref{module_call_end} create a call to that function and insert it into the IR.}
\label{ir-idiosyncratic}
\end{figure*}
\end{itemize}

WALA uses this machinery for multiple languages,  JavaScript from
Rhino and Java source from Eclipse. It has been public for several years now, used
by researchers for a range of work, such as hybrid app
analysis\cite{DBLP:conf/kbse/LeeDR16}, correlation
tracking\cite{DBLP:conf/ecoop/SridharanDCST12}, a special form of
context sensitivity, tunable analysis techniques~\cite{7372042}, and approximate analysis to handle real-world JavaScript code\cite{DBLP:conf/icse/FeldthausSSDT13}.  It has been robust and scalable enough to support products, and the Python analysis for machine learning leverages that technology. 

\subsection{IR extensions}

Beyond customizing IR generation from the source code, WALA also
provides ways to create synthetic IR to represent aspects of the code
not directly present in the source. This machinery involves writing
explicit models of functions as IR using something analogous to an
assembler for IR.  There are two way of constructing such models that
we employ in this work, both of which have been used extensively in
the past to model Java and JavaScript.

\paragraph{Programmatic} \label{python-ir} Creating IR with analysis code is the most
flexible way to create IR, and WALA provide many ways to integrate
such IR into analysis.  Python has relatively straightforward
semantics in many cases, and needs relatively little such modeling;
however, the precise semantics of method calls are a little tricky due
to how {\tt self} is handled.  In effect, all normal method
calls in Python act like calls to a normal function closed over the
{\tt self} object.  Consider the following snippet of
code:

\begin{Verbatim}[commandchars=\@\|\^,numbers=left,numbersep=6pt]
class Foo(object):
  def foo(self, a):
    return self.contents+a

x = Foo()
y = x.foo  

x.foo(3) @label|normal_foo_call^
Foo.foo(x, 3) @label|function_foo_call^
y(3) @label|closure_foo_call^

x.foo = Foo.foo
x.foo(x, 3) @label|mutated_foo_call^
\end{Verbatim}

 The calls at lines~\ref{normal_foo_call},~\ref{function_foo_call},~\ref{closure_foo_call} and~\ref{mutated_foo_call} are all equivalent, but the latter are
unusual in object-oriented languages.  When the method {\tt foo} is
read from the class {\tt Foo}, the result is an unbound method that
needs to be called with an explicit {\tt self} parameter, but when it
is read from the object {\tt x}, the result is a method bound to that
object.  That is like a closure, since the call on
line~\ref{closure_foo_call} carries the {\tt self} object along with
the method.  And finally observe that methods are writable just like
normal fields, resulting in the call at line~\ref{mutated_foo_call}.

We model this with synthetic IR that, for every class, creates a
constructor method that initializes each method field with a trampoline
that captures its self object to pass as part of a call to the real
method from the
trampoline.  This has the effect of making calls like
line~\ref{normal_foo_call} and~\ref{closure_foo_call} work and allows
the field to be mutated as required.

\paragraph{XML}\label{tf-model} We choose to model popular machine
learning frameworks in order to explicitly capture their semantic properties, using an XML representation of the
IR.  We focus in Figures~\ref{wala-tensorflow-model-init} and~\ref{wala-tensorflow-model} on aspects of the popular Tensorflow framework from our example.  Figure~\ref{wala-tensorflow-model-init} shows the model of importing Tensorflow, which creates the key objects in the Tensorflow model.  Figure~\ref{wala-tensorflow-model} shows models of functions of the Tensorflow objects.  For example, on line~\ref{create_estimator}, the estimator object is created, which we model in two pieces:  The Estimator object is allocated at line~\ref{make_Estimator} of Figure~\ref{wala-tensorflow-model-init} and the call is modeled by the IR at line~\ref{Estimator_ctor} of Figure~\ref{wala-tensorflow-model}.  The {\tt reshape} call at line~\ref{reshape} of Figure~\ref{mnist-example} is analogous, with the reshape object defined at ~\ref{reshape_alloc} in Figure~\ref{wala-tensorflow-model-init}.

 \begin{figure}[p]
  \centering
 {\scriptsize
\begin{Verbatim}[commandchars=\@\|\^,numbers=left,numbersep=6pt]
    <class name="tensorflow" allocatable="true">
      <method name="import"
	      static="true"
	      descriptor="()Ltensorflow;">
        <new def="x" class="Ltensorflow"/>

        <new def="estimator" class="Lobject"/>
	<putfield class="LRoot"
                  field="estimator"
                  fieldType="LRoot"
                  ref="x"
                  value="estimator"/>

	<new def="Estimator" @label|make_Estimator^
                  class="Ltensorflow/estimator/Estimator"/>
	<putfield class="LRoot"
                  field="Estimator"
                  fieldType="LRoot"
                  ref="estimator"
                  value="Estimator"/>

	<new def="inputs" class="Lobject"/>
	<putfield class="LRoot"
                  field="inputs"
                  fieldType="LRoot"
                  ref="estimator"
                  value="inputs"/>

	<new def="numpy_input_fn" 
                  class="Ltensorflow/estimator/numpy_input_fn"/>
	<putfield class="LRoot"
                  field="numpy_input_fn"
                  fieldType="LRoot"
                  ref="inputs"
                  value="numpy_input_fn"/>

	<new def="reshape" class="Ltensorflow/functions/reshape"/> @label|reshape_alloc^
	<putfield class="LRoot"
                  field="reshape"
                  fieldType="LRoot"
                  ref="x"
                  value="reshape"/>

	<return value="x"/>
      </method>
    </class>
\end{Verbatim}
}
\caption{Excerpt of TensorFlow model in WALA.  This excerpt represents the initialization of the objects constituting the model of TensorFlow used by WALA.  The TensorFlow object itself is created, its {\tt Estimator} and {\tt reshape} elements are created, and these are linked with fields that represent the TensorFlow API exposed to applications.  The elements of the model are a direct representation of WALA IR in an XML format. The method {\tt import} is invoked by the translation of import statements in Python.  The {\tt new} and {\tt putfield} elements represent WALA IR object creations and field writes respectively.  The {\tt def} properties are names of local variables defined, and {\tt value} and {\tt ref} represent reads of those variables corresponding to values and objects of field writes.}
\label{wala-tensorflow-model-init}
\end{figure}

\begin{figure}[p]
  \centering
 {\scriptsize
\begin{Verbatim}[commandchars=\@\|\^,numbers=left,numbersep=6pt]
<package name="tensorflow/estimator">
 <class name="Estimator" allocatable="true"> @label|estimator_class^
   <method name="do" descriptor="()LRoot;" numArgs="2"> @label|function_body_name^ @label|Estimator_ctor^
     <new def="x" class="Ltensorflow/estimator/train/train"/>
     <putfield class="LRoot"
               field="train"
               fieldType="LRoot"
               ref="arg0"
               value="x"/>
     <putfield class="LRoot" @label|store_function^
               field="$callback"
               fieldType="LRoot"
               ref="x"
               value="2"/>
     <return value="arg0"/>
   </method>
 </class>

 <class name="numpy_input_fn" allocatable="true">
   <method name="do" descriptor="()LRoot;" numArgs="3">
     <new def="x" class="Lobject"/>
     <putfield class="LRoot"
               field="data"
               fieldType="LRoot"
               ref="x"
               value="2"/>
     <putfield class="LRoot"
               field="labels"
               fieldType="LRoot"
               ref="x"
               value="3"/>
     <return value="x"/>
   </method>
 </class>
/package>

<package name="tensorflow/estimator/train">
 <class name="train" allocatable="true">
   <method name="do" descriptor="()LRoot;" numArgs="3">
     <getfield class="LRoot" @label|read_function^
               field="$callback"
               fieldType="LRoot"
               ref="arg0"
               def="x"/>
     <call class="LRoot"
   	name="do"
   	descriptor="()LRoot;"
   	type="virtual"
   	arg0="x"
   	arg1="2"
   	arg2="3"
   	numArgs="3"
   	def="v"/>
     <return value="v"/>
   </method>
 </class>
</package>
\end{Verbatim}
}
\caption{Excerpt of Tensorflow model in WALA.  The functions of the Tensorflow.  The method name and descriptor at line~\ref{function_body_name} is how Python functions are represented in WALA IR, so this portion of the model describes the functions of {\tt Estimator}, {\tt numpy\_input\_fn} and {\tt train}.  {\tt Estimator} is called with 2 arguments, the Estimator object itself and a callback function; it allocates and stores the {\tt train} object for the Estimator, and stores the callback function in it.  The {\tt numpy\_input\_fn} function stores the input data, which is its argument.  The {\tt train} function invokes the callback given to the Estimator.}
\label{wala-tensorflow-model}
\end{figure}

\section{Types}
\label{sec:types}

\begin{definition}[Types for ML]
  \label{def:types}
  \[
    \begin{array}{lrcl}
      (\mbox{Python Type})
      &\pi& ::= & \trec{f_1 : \pi_1 , \ldots , f_n:\pi_n}\\
      &&\mid&(f_1 : \pi_1, \ldots, f_n:\pi_n)\to\pi\\
      &&\mid& \tau\mid l\mid\ttop\\
      (\mbox{Tensor Type})&\tau & ::=&\ttensor{d_1 , \ldots , d_n}{\pi}\\
      (\mbox{Dimension})&d & ::=& l \mid n \mid l(n) | d*d \\
    \end{array}
  \]
  
  where $f$ are record and parameter names (strings), $l$ are labels (strings), and $n$ are non-negative integers.

  We ignore the ordering of fields in a record and consider the $*$ operator to be associative.
\end{definition}

Defintion~\ref{def:types} presents a simple type system for tracking tensors in a Python program.  Python types,
$\pi$, track values having tensor type.  Additionally,
since records may contain tensors, and functions may consume and
return tensors, these are also modelled. These types can be easily
extended to handle other structures in python (e.g. lists).  We are
deliberately simplifying these features to emphasize the contribution
of this paper, tensor types.

To fucs on tensors, it also includes the
top type, \ttop, which to indicate irrelevant types.
It also allows labels to identify the category/type
of the actual data.
By convention, function and record arguments of type $\ttop$ are
omitted.

Tensor types $\tau$, track the contents and
dimensions of a tensor/matrix/vector.  The second part of a tensor type
describes the data that it contains.  The first part is a
list of the dimensions.  This list notation denotes nested tensor types, each with a single dimension.  In
particular, these are equivalent:
\[\ttensor{d_1 , d_2 , \ldots ,  d_n}{\pi} = \ttensor{d_1}{\ttensor{d_2 , \ldots , d_n}{\pi}}\]

\subsection{Example Walkthrough}
\label{sec:types-example}

We illustrate the types using Figure~\ref{mnist-example}; comments
(starting with \verb|##|) specify type information for function
definitions and relevant variable assignments. 

MNIST data is read on line~\ref{mnist_read}, using
\verb|read_data_sets|; the returned data is modelled with type:
\begin{align*}
\{\tfield{train}:\trec{\tfield{images} : \ttensor{\tlabel{batch},
      \tlabel{y}(28)*\tlabel{x}(28)}{\tlabel{channel}}},\\
    \tfield{test}:\trec{\tfield{images} :  \ttensor{\tlabel{batch},
      \tlabel{y}(28)*\tlabel{x}(28)}{\tlabel{channel}}}\}  
\end{align*}It has (at least) two fields, \tfield{train}
and \tfield{test}, each of which contain a nested structure with an \tfield{images}
field that contains a tensor.  This tensor has two dimensions: the
outer dimension denotes images, and the inner dimension is a
\tlabel{channel} (number) for each pixel in the image.  The inner
dimension has size $784$, which represents a flattened $28x28$
matrix of channels.  The dimension type is uses a combined dimension,
$28*28$, to capture this; using labels, it captures their order: the first
part of this dimension represents the height and the second the
width. The type $\tlabel{y}(28)*\tlabel{x}(28)$ preserves this.

Around Line~\ref{heap_write_images},
the mnist \tfield{train}ing data, which has the type
\[\trec{\tfield{images} : \ttensor{\tlabel{batch}, \tlabel{y}(28) *
    \tlabel{x}(28)}{\tlabel{channel}}}\] and the training labels, with
type $\ttensor{[10}{\tlabel{label]}}$ are used to create the input for
the estimator.

These arguments will be passed into the estimator model, which, as
shown on Line~\ref{create_estimator}, is \verb|model_fn| method
defined on Line~\ref{def_model_fn}.  This call is validated by our
system, as the types of the arguments agree with their expected
types, shown in the comments.  The first thing this method does is
call the \verb|conv_net| method defined on Line~\ref{def_conv_net}.
Again, it is easy to see that the types of the data passed in to the
method are compatible with the parameters' expected types.

Finally, we the \verb|conv_net| method, which
creates the neural network.  On Line~\ref{heap_read_images}, the
method first gets the tensor with the actual image data, of type
$\ttensor{\tlabel{batch}, \tlabel{y}(28) *
  \tlabel{x}(28)}{\tlabel{channel}}$.  It then preforms a reshape
operation, turning it into a 4-dimensional vector of channels, with
type
$$\ttensor{\tlabel{batch}, \tlabel{y}(28),
  \tlabel{x}(28),1}{\tlabel{channel}}$$ Note how the (common)
free-form comments about the input and output dimensionality are
captured by the types. 

This \verb|reshape| operation highlights the advantage of keeping
structured combined dimensional information.  In the serialized mnist
data, each image is represented by a flattened vector of $784$.
However, logically, it is a $28x28$
matrix of channels.  That information allows checking that requested
reshape is reasonable. 
Had the code tried to reshape the data into a $56x14$ matrix,that is
not a valid factorization.  

Next, on Line~\ref{conv1}, the reshaped tensor is passed into the
\verb|conv2d| method.  This method requires that the input tensor has
four dimensions, a constraint that our type for the input easily be
seen to satisfy.  It also requires that the contents of the dimension
be a number.  In our example, the \tlabel{channel} label does indeed
represent a numeric value.
Finally, it requires that the middle two
dimensions represent \tlabel{height} and \tlabel{width} (or \tlabel{y}
and \tlabel{x}, or similar).  This can be checked by observing the
labels of the dimensions of the input tensor.

Checking the meaning of a dimension,
provides a new way to find common bugs.   In particular, it would be
easy to accidentally swap the \tlabel{y} and \tlabel{x} dimensions in
our example.  To prevent this kind of bug, multiple machine learning
experts have independently stumbled upon a common technique: ensuring
that dimensions are all of different sizes.  In our example, this
could be done by artificially changing the image size to,
e.g. $28x29$.  The type system presented here allows this information
to instead be captured symbolically in the type system.

Continuing on in the \verb|conv_net| method, Line~\ref{pool1}
downsamples the given tensor, resulting in a tensor of type\\ 
$\ttensor{\tlabel{batch}, \tlabel{y}(14),
  \tlabel{x}(14),1}{\tlabel{channel}}$.  Line~\ref{pool2} downsamples
it again, and Line~\ref{flatten1} flattens the resulting tensor,
yielding a tensor with type $\ttensor{\tlabel{batch}, \tlabel{y}(7)*
  \tlabel{x}(7)}{\tlabel{channel}}$.



\section{Analyzing Tensors}
\label{sec:dataflow}

 WALA produces a dataflow graph as part of pointer analysis and call graph construction.  The dataflow graph summarizes the flow of objects and values in the program; this graph is an abstraction of possible program behavior, and is defined as follows:
\begin{definition}[Dataflow Graph]
A {\em dataflow graph} $\mathfrak{G}=\left<V,S,\prec\right>$ where $V$ is the set of program variables, $S(v)$ is the set of objects or values possibly held by $v \in V$ and $x \prec y$ iff there is potential dataflow from $y \in V$ to $x \in V$, e.g. through an assignment or function call.
\end{definition}

Given a dataflow graph $\mathfrak{G}$, we defined a {\em tensor estimate} $T(v)$ as the set of possible tensor types that tensors held by $v$ may have.  We use $\mathfrak{I}$ to denote the declared tensor type of input; {\tt reshape} takes a tensor and object representing a tensor type (a list of numbers in Python) and produces a reshaped tensor if the input tensor can be reshaped to the given size, denoted $\doteq$.  This is implemented directly using WALA dataflow analysis, and is defined as follows:
\begin{definition}[Tensor Estimate]
Given a dataflow graph $\mathfrak{G}$, a {\em tensor estimate} $\mathfrak{T}(\mathfrak{G})=\left<T\right>$ defines the set of tensor types a variable may take on.  The  is defined as either the given input type, dataflow in the program or the result of a {\tt reshape} or other Tensorflow API:
$$T(y) \subseteq \left\{
\begin{array}{ll}
\left\{\mathfrak{I}\right\} & \hbox{$y$ is input} \\
T(x) & y \prec x\\
z & y \prec {\tt reshape(x, z)} \wedge \exists_{z_i \in S(z)} T(x) \doteq z_i\\
\ldots & y \prec \hbox{other Tensorflow APIs}
\end{array}\right. $$
\end{definition}

\section{Results}

 We evaluated our prototype implementation by analyzing 5 Tensorflow examples based on MNIST and 1 Tensorflow-based code that analyzes neural imaging data.  The codes are chosen because tutorials based on MNIST are common, and neuroscience is an area of focus for this project. Use of Tensorflow to analyze neural imaging data is an emerging area, and we know of few public codes that we can test.  Our model of Tensorflow is not complete at this point, and so we focus on a subset of its API to illustrate our analysis.
\begin{description}
\item[{\tt{reshape(tensor, shape)}}] is part of the API illustrated in Figure~\ref{mnist-example}; it reshapes a tensor to the dimensions specified in its second argument, which is a Python list.  Our analysis checks that the tensor argument is the right size to be transformed into the desired shape.
\item[{\tt{conv2d(tensor, ...)}}] is part of the API illustrated in Figure~\ref{mnist-example}; it performs a 2D convolution on images, and it requires the images to be in a format of height, width, layer, where layers represent different colors.  Our analysis checks that the argument has that structure.
\item[{\tt{conv3d(tensor, ...)}}] is analogous to {\tt conv2d}, but for 3D images.  Since neural images are fMRI scans, they are 3D images of the brain, and hence {\tt conv3d} is used.
\item[{\tt{placeholder(shape)}}] represents an unspecified tensor.  It is used for operations that are later applied to data; for instance, it can be used to create a neural network which is then trained by feeding multiple dataset batches into the network.  This is called {\tt place} in Table~\ref{results}
\end{description}

The six programs on which we evaluate are the following:
\begin{description}
\item[conv\_network\footnotemark]\footnotetext{\url{https://github.com/aymericdamien/TensorFlow-Examples/blob/master/examples/3_NeuralNetworks/convolutional_network.py}} builds and trains a convolutional neural network to classify MNIST data.  Reshaping and convolving the images is part of the network.
\item[mnist\_deep\footnotemark]\footnotetext{\url{https://github.com/tensorflow/tensorflow/blob/master/tensorflow/examples/tutorials/mnist/mnist_deep.py}} is an  MNIST classifier using convolutional layers from the Tensorflow github site. \item[mnist\_max\footnotemark]\footnotetext{\url{https://github.com/tensorflow/tensorflow/blob/master/tensorflow/examples/tutorials/mnist/mnist_softmax.py}} is a simple MNIST classifier from the Tensorflow github site.  It does not reshape, but it uses placeholders to allow training in batches.
\item[mnist\_max\_xla\footnotemark]\footnotetext{\url{https://github.com/tensorflow/tensorflow/blob/master/tensorflow/examples/tutorials/mnist/mnist_softmax_xla.py}} is a simple MNIST classifier from the Tensorflow github site.  It has options for optimization, but analysis is similar to {\tt mnist\_softmax}.
\item[mnist\_sum\footnotemark]\footnotetext{\url{https://github.com/tensorflow/tensorflow/blob/master/tensorflow/examples/tutorials/mnist/mnist_with_summaries.py}} is a simple MNIST classifier from the Tensorflow github site.  It reshapes the input images to generate a viewable summary.
\item[neuroimage\footnotemark]\footnotetext{\url{https://raw.githubusercontent.com/corticometrics/neuroimage-tensorflow/master/train.py}} classifies 3D images of brain data.  It resembles MNIST as image classification, but the images are 3D and so use different Tensorflow APIs.
\end{description}

 Although our implementation is still preliminary, it is sufficient to parse these programs and to analyze them, at least partially.  Table~\ref{results} shows the APIs that were analyzed in each program; for each API, our analysis was able to verify that these programs used them in correct ways.  Given the preliminary nature of our implementation so far, we make no claims about false negatives; that is, there may be uses of the these constructs that are not analyzed fully in addition to the ones we did analyze.  The \Checkmark notation in Table~\ref{results} denotes constructs that we found in the codes and were able to verify were used properly in at least some cases; our analysis so far found no false positives in these programs.  The \XSolidBrush denotes that we verified that the program does not contain the construct; thus, since there are no empty spots in Table~\ref{results}, any false negatives can only be from incomplete analyses of constructs that we found.  Furthermore, the analysis of each program took just a few seconds on a normal laptop. All of these results were generated by our code on Github\footnote{\url{https://github.com/wala/ML}} and the analysis can be seen in our tests on Travis CI\footnote{\url{https://travis-ci.org/wala/ML}}.

\begin{table}[htb]
\begin{tabular}{l|c|c|c|c}
program&reshape&conv2d&conv3d&place\\ \hline
conv\_network & \Checkmark & \Checkmark & \XSolidBrush & \XSolidBrush \\
mnist\_deep & \Checkmark & \Checkmark & \XSolidBrush & \Checkmark\\
mnist\_max & \XSolidBrush & \XSolidBrush & \XSolidBrush & \Checkmark \\
mnist\_max\_xla & \XSolidBrush & \XSolidBrush & \XSolidBrush & \Checkmark \\
mnist\_sum & \Checkmark & \XSolidBrush & \XSolidBrush &\Checkmark \\
neuroimage & \Checkmark & \XSolidBrush& \Checkmark & \XSolidBrush\\
\end{tabular}
\caption{Tensorflow APIs analyzed}
\label{results}
\end{table}

\section{Related Work}

 There has been some work in program analysis for Python, and we'll discuss the most-related tools and summarize the rest of the work.
\paragraph{Python Taint}\cite{pyt} is a static analysis tool for detecting security vulnerabilities in Python.  It uses standard dataflow techniques, and can do some interprocedural analysis.  However, its interprocedural analysis is limited: it looks for a definition of a function for a call using its name, rather than handling function pointers and object semantics as required by the language and needed for even our simple example.  These sorts of features depend on the sort of rich analysis infrastructure provided by WALA. 

The rest of the work breaks down into two broad classes, code quality checkers and dynamic analysis.
Code quality checkers are static analyzers for code quality metrics or lint tools; examples are Pylint\cite{Pylint}, pycodestyle\cite{pycodestyle}, pyflakes\cite{pyflakes}, Flake8\cite{Flake8}, QuantifiedCode CE\cite{quantified_code}, pydocstyle\cite{pydocstyle}. There are so many static analysis tools for code quality checking, that there are sites like Awesome Static Analysis\cite{awesome_static_analysis} which keep up to date with the latest developments in the tools, for Python and for many other languages. Many of these tools are conveniently packaged in a wrapper like Prospector\cite{Prospector}, which combine several tools. Prospector gives the ability to turn particular tools, features, or behaviors on and off as needed, and starts with easy-to-use defaults. Others include AST-based tools like jedi\cite{jedi}, bandit\cite{bandit} and mccabe\cite{mccabe}.  These tools are all local analyses, for instance, mccabe focuses on the syntactic code complexity of single functions and many focus on code style issues.  These tools are not for whole-program static analysis based on dataflow, as WALA is.

Dynamic analyses do sometimes build call graphs, of which a notable example is simply called ``Python CallGraph''\cite{pycg}. Another dynamic analysis tool, PyChecker\cite{PyChecker}, does things one would expect a compiler to do for many other languages. For example, it ensures variables are set before they are used, it checks that class methods are used properly, and ensures that the number of arguments passed into a function is correct. These tools are very different from WALA, as our goal is to statically approximate all possible runs, rather than dynamic tools that focus on one or a few executions.  There is a library-based taint analysis tool for Python \cite{10.1007/978-3-642-27937-9_15}; unlike many other dynamic analysis tools, it is implemented purely in Python, meaning that it can be used without any modifications to the interpreter, and adapted easily.

There has been other work in types for tensors~\cite{DBLP:journals/corr/abs-1710-06892} from the SCALA'17 workshop.  This work uses heterogeneous lists to represent the semantics of dimensions of tensors; however, this work is in Scala and so it is not integrated as ours is to to e.g. check reshape operations in Python.

There is also a lot of interesting work in the area of using machine learning to do analysis of code \cite{DBLP:journals/corr/abs-1709-06182}. However, this work remains technically distinct, in that it uses machine learning to do analysis rather than applying code analysis to aid machine learning.

\section{Conclusions}

 We have presented our application of WALA to analyzing the behavior of tensors in Tensorflow programs, using types to track their shapes.  This involved adapting WALA to Python, which we believe is the first application of traditional whole-program analysis technology to Python.

\bibliography{refs}

\end{document}